\documentclass[letterpaper,twocolumn,10pt]{article}

\usepackage{settings}

\usepackage{enumitem}
\usepackage{multirow}   
\usepackage{booktabs}    
\usepackage{mathtools}   
\usepackage{enumitem}    
\usepackage{graphicx}     
\usepackage{cleveref}   
\usepackage{xspace}       
\usepackage{float}   
\usepackage{makecell}     
\usepackage[table]{xcolor}
\usepackage{soul}

\usepackage{etoolbox}
\AtBeginEnvironment{quote}{\setlength{\topsep}{2pt}}

\usepackage{tikz}
\usepackage{amsmath}

\usepackage{filecontents}

\begin{document}

\date{}

\title{\Large \bf Investigating In-Context Privacy Learning by Integrating User-Facing Privacy Tools into Conversational Agents}

\def\plainauthor{Author name(s) for PDF metadata. Don't forget to anonymize for submission!}

\author{
{\rm Mohammad Hadi Nezhad}\\
University of Massachusetts Amherst\\
mhadinezhad@cs.umass.edu
\and
{\rm Francisco Enrique Vicente Castro}\\
New York University
\and
{\rm Ivon Arroyo}\\
University of Massachusetts Amherst
}

\maketitle

\begin{abstract}
Supporting users in protecting sensitive information when using conversational agents (CAs) is crucial, as users may undervalue privacy protection due to outdated, partial, or inaccurate knowledge about privacy in CAs. 
Although privacy knowledge can be developed through standalone resources, it may not readily translate into practice and may remain detached from real-time contexts of use.
In this study, we investigate in-context, experiential learning by examining how interactions with privacy tools during chatbot use enhance users' privacy learning. 
We also explore interface design features that facilitate engagement with these tools and learning about privacy by simulating ChatGPT's interface which we integrated with a just-in-time privacy notice panel. 
The panel intercepts messages containing sensitive information, warns users about potential sensitivity, offers protective actions, and provides FAQs about privacy in CAs.
Participants used versions of the chatbot with and without the privacy panel across two task sessions designed to approximate realistic chatbot use.
We qualitatively analyzed participants' pre- and post-test survey responses and think-aloud transcripts and describe findings related to (a) participants' perceptions of privacy before and after the task sessions and (b) interface design features that supported or hindered user-led protection of sensitive information.
Finally, we discuss future directions for designing user-facing privacy tools in CAs that promote privacy learning and user engagement in protecting privacy in CAs.
\end{abstract}

\section{Introduction}
Protecting privacy during conversational agent (hereafter CAs or chatbots) use requires users to continuously assess what information---about themselves or others---\emph{``is appropriate, or fitting, to reveal in a particular context''} \cite{nissenbaum-2004}. 
However, users often disclose sensitive information that may not be necessary for completing tasks \cite{zhang-2024,malki-2025-hoovered,mireshghallah-2024}.
Even when such information is relevant, disclosure decisions are not always made by a careful weighing of immediate benefits against potential future privacy risks. 
Examples include treating CAs as therapists and sharing emotional or behavioral details, providing medical or fitness information to receive advice, or uploading lengthy documents (e.g., emails or contracts) for summarization or search \cite{zhang-2024,mireshghallah-2024}. 
In pursuing these goals, users may overlook the sensitivity of the information they are disclosing, be unaware of available actions to protect their information, or find such protective actions time-consuming or inconvenient. 
As a result, they may undervalue privacy protection and its importance during CA interactions. 
This makes it essential to support users in more context-aware and privacy-conscious decision-making. 

Given these concerns, it is essential to support different user groups in developing a stronger understanding of privacy in the context of CAs, as supporting them can promote greater and more meaningful engagement with privacy-conscious behaviors \cite{Zhou-2025-rescriber,clear-2025}. 
While such understanding can be fostered through standalone learning resources (e.g., educational tools or academic programs) \cite{feffer-2023,hadi-2025-embedding}, it may not easily translate into practice, as it is often detached from the real-time contexts of use in which privacy decisions must be made (e.g., what information to protect and how, depending on the tasks and constraints including time and effort) and where timely awareness and support are most critical.
\textit{We thus explore how exposing users to privacy tools during realistic chatbot use scenarios can support experiential, in-context learning about privacy, while also exploring user interface and experience (UI/UX) design features that encourage user engagement with privacy protections and facilitate learning.} 

We first examine key aspects of privacy perceptions among undergraduate and master's students in a Computer Science (CS) program in the US (participant details in Section \ref{section_participants}). 
We then analyze how these perceptions evolve through interactions with a just-in-time \textit{privacy notice panel} (Section \ref{sec_interface}) that embeds learning opportunities directly within CA interfaces. 
To this end, we integrated a simulated ChatGPT interface with our privacy panel that intercepts interactions immediately after users attempt to submit a message containing sensitive information. 
Our panel (1) detects and informs users about potentially sensitive information in their messages (e.g., names, physical addresses); (2) offers features for applying anonymization strategies (retracting, faking, generalizing); (3) surfaces two built-in ChatGPT privacy controls (disabling memory, opting out of data sharing); and (4) includes FAQs that describe how to think about sensitive information and factors to consider when making disclosure decisions.

We analyzed our panel's effectiveness through a five-phase study (Section \ref{sec_procedure}) with ten participants: (1) a \textit{pre-test survey} examining baseline privacy perceptions; (2–3) two \textit{study sessions to complete tasks} using the simulated chatbot with and without our privacy panel; (4) an \textit{immediate post-test survey} collecting reflections on the task sessions; and (5) a \textit{delayed post-test survey} to examine changes in privacy perceptions.
Our task and study design closely approximate realistic chatbot use for our participant group by incorporating a range of real-world use scenarios involving sensitive information (e.g., writing emails, searching documents), withholding the study's privacy goals until after participation to reduce bias, and allowing participants to freely direct their own interactions (e.g., prompting, disclosure decisions).
Thus, we examine the following research question (RQ):
\begin{enumerate}[label=\textbf{RQ\arabic*.}, leftmargin=2.7em]
    \item \textit{Before} and \textit{after} completing tasks using the chatbot, what did students think about---
    \begin{enumerate}[label=\textbf{\arabic{enumi}.\arabic*.}, leftmargin=1.5em]
        \item what they consider as sensitive information?
        \item the importance of protecting sensitive information when using chatbots?
        \item what they can do to protect sensitive information when using ChatGPT?
    \end{enumerate}
\end{enumerate}

Given the critical role of UI/UX design in users' engagement with and adoption of interactive tools, there are gaps and opportunities to better understand how interface design decisions support or hinder the protection of sensitive information when using CAs.
Thus, we adopt an exploratory approach to examine such design features in our privacy panel by analyzing students' think-aloud transcripts and interface interactions from task sessions and responses to open-ended survey questions.  
Therefore, we examine the following RQ:
\begin{enumerate}[label=\textbf{RQ2.}, leftmargin=2.7em]
    \item What interface design features of the privacy panel support or hinder the protection of sensitive information?
\end{enumerate}

Through a detailed qualitative analysis of data (Section \ref{sec:analysis}), we present findings on how our participants conceptualized sensitive information (Section \ref{results_rq1.1}), viewed the importance of protecting sensitive information during CA use (Section \ref{results_rq1.2}), and what protective actions they thought they can take (Section \ref{results_rq1.3}). 
We also report findings on interface design decisions that supported or hindered users' protection of sensitive information when interacting with our privacy panel (Section \ref{results_rq2}).
We then discuss opportunities for future work on enhancing in-context support and experiential privacy learning (Section~\ref{sec_dis_experiential}), designing UI/UX to promote engagement and learning in user-facing privacy tools (Section~\ref{sec_dis_uiux}), supporting users in navigating privacy trade-offs (Section~\ref{sec_dis_tradeoffs}), and identifying learning opportunities for students (Section~\ref{sec_dis_learning_opp}).

\section{Background and Related Work}
\label{sec:rw}
\subsection{User Knowledge and Agency Over Privacy When Using Conversational Agents (CA)}
Users need to understand and consider various aspects of privacy in their decision-making to meaningfully engage in privacy-protective behaviors in CAs use. 
Decisions about what to disclose or withhold may be shaped by several factors, including individual preferences (e.g., sensitivity of financial information may differ across individuals \cite{Belen-Saglam-2022}), the features of the technology (e.g., privacy policies, security safeguards), and the perceived trustworthiness of the technology providers in protecting user privacy \cite{lau-2018-alexa,Song-2025,zhang-2024,hadinezhad-2026}.
However, users’ awareness and understanding of these factors may not be updated by, for example, current data practices.
For instance, research shows that users often rely on partial, simplified, or inaccurate mental models of how LLM-based CAs process their data \cite{zhang-2024}. 
They may not recognize what types of information can be used to identify individuals \cite{Song-2025,hadinezhad-2026}, or may underestimate the potentials of these tools in exacerbating the expected consequences of data leakage---for themselves and others---through capabilities such as real-time, detailed profiling that can heighten the risks of data access and misuse \cite{lee-2024-deepfakes,kelley-2023}.
Moreover, even when users intend to protect sensitive information, they may be unaware of available privacy-protective actions, whether built into the tool (e.g., ChatGPT's option to opt-out of content sharing for model training) or actions they can take themselves to address the tool's privacy limitations (e.g., anonymizing sensitive data before submission) \cite{zhang-2024,hadinezhad-2026}. 
These gaps in knowledge and awareness can lead users to overlook or undervalue privacy protection, especially when weighing trade-offs such as deciding whether to disclose more sensitive information to improve chatbot performance or withhold it to safeguard privacy.

Research has suggested that privacy perceptions (e.g., regarding information sensitivity) can vary across people based on factors such as age, gender, education, and geographic location \cite{malki-2025-hoovered,Belen-Saglam-2022,Tifferet-2019,Engstrom-2023,Li-2017-cross-cultural}. 
This highlights the need to examine privacy perceptions across different user groups, avoid overgeneralizing findings, and tailor support to the specific needs of each group.
For example, a study with UK-based CA users aged 18 and over found that participants lacked reliable strategies for protecting privacy, had difficulty understanding privacy features and their outcomes, and showed low baseline awareness of privacy risks \cite{malki-2025-hoovered}. 
In another study, researchers investigated how users of women's period and fertility tracking apps define and understand personally identifiable information (PII), revealing mismatches between users' perceptions and regulations regarding PII protections \cite{Song-2025}.

In this study, we examined the privacy perceptions of Computer Science (CS) undergraduate and master's students at a US university (details in Section \ref{section_participants}). 
Specifically, we explored how they conceptualize sensitive information, how important it is to protect such information during CA use, and what protective actions they think they can take. 
We also investigate how their views changed after interacting with our privacy panel while completing realistic chatbot tasks.

\subsection{User-Facing Privacy Tools for Enhancing Knowledge During CA Use}
Interactive systems can support knowledge development by embedding learning opportunities directly within users' workflows. 
This can enable learning to occur in context and through experience---e.g., through real tasks, constraints, and goals (see also \textit{Experiential Learning} \cite{illeris-2007, kolb-2014}).
Accordingly, integrating user-facing privacy tools into CAs can support users in engaging with privacy-protective practices, such as recognizing sensitive information and protective approaches \cite{Zhou-2025-rescriber,hadinezhad-2026}. 
When these tools are integrated into the interaction context, they can help highlight the necessity of protections (e.g., observing the extent of sensitive info being disclosed, reflecting whether protections impact chatbot performance).
This, in turn, can incentivize users to take protective actions, engage more deeply with thinking and learning about privacy in CAs, and apply this acquired knowledge in their future interactions.

Scholars have developed user-facing privacy tools for CAs to promote privacy awareness and encourage protective actions. 
For example, \textit{Clear} \cite{clear-2025} is a just-in-time interface that automatically identifies sensitive information in user messages and informs users of relevant privacy policies and potential disclosure risks. 
\textit{Rescriber} \cite{Zhou-2025-rescriber} assists users in detecting and sanitizing sensitive content in their prompts. 
\textit{Casper} \cite{chong-2024-casper} automatically anonymizes PII and notifies users about sensitive topics in their messages.
However, these studies did not systematically analyze changes in users' lasting privacy knowledge, focusing instead on short-term awareness. 
Only the \textit{Rescriber} study reported insights from short-term reflections, suggesting that participants had perceptions of learning.

In this study, students interact with our privacy panel, which provides learning opportunities through identifying and informing users about certain types of potentially sensitive information in their messages, offering anonymization strategies (retracting, faking, and generalizing), surfacing two built-in chatbot privacy controls (opting out of content sharing for model training, disabling memory), and presenting a set of FAQs about privacy in CAs.
We assess changes in participants' lasting privacy knowledge (i.e., ability to recall and explain) by administering pre- and post-test surveys at least one week before and after the task sessions.

\subsection{UI/UX Design for Engagement with Privacy Tools}
The design of the UI/UX plays a critical role in the adoption and effective engagement with user-facing tools. 
In interactive systems, users' choices are often shaped by factors including how options are presented, when they are surfaced, what alternative choices are offered, and which default options are provided (particularly given that default options tend to be selected more frequently) \cite{johnson-2012, thaler-2009, martins-2021}. 
For example, prior work shows that certain entry points to privacy settings can improve their discoverability, findability, and perceived usability \cite{Im-2023-less}.
Similarly, privacy notices that are presented at the moment of decision-making and tailored to users' context have been found to more effectively support privacy-protective choices \cite{Schaub-2015, Kelley-2013, clear-2025}.
Scholars have raised related concerns in the context of privacy tools in CAs.
For example, ChatGPT uses user content for model training by default and although an opt-out option exists, studies show that many users are unaware of it or do not understand how it works \cite{zhang-2024,hadinezhad-2026}.
The convenience of user actions in the tool also plays a key role in disclosure and protection behaviors. 
For instance, chatbot users are more likely to share information when disclosure is easily afforded by the interface (e.g., features for uploading lengthy documents) \cite{zhang-2024}. 
On the privacy-tool side, users value designs that simplify and streamline sanitization efforts (e.g., bulk sanitization), while also reporting challenges in understanding certain sanitization strategies (e.g., abstraction) \cite{Zhou-2025-rescriber,hadinezhad-2026}.
This highlights a need to further examine UI/UX designs that support engagement with privacy tools in CAs.
In this work, we adopt an exploratory approach to identify features in the design of our privacy notice panel that supported or hindered the protection of sensitive information.

\section{Methods}
In this section, we describe our ChatGPT interface simulation and privacy notice panel design (Section \ref{sec_interface}), the study procedure (Section \ref{sec_procedure}), participants (Section \ref{section_participants}), and data collection and analysis approach (Section \ref{sec:analysis}).

\subsection{Chatbot and Privacy Notice Panel Design}
\label{sec_interface}
\begin{figure*}[t]
  \centering
  \includegraphics[width=\linewidth]{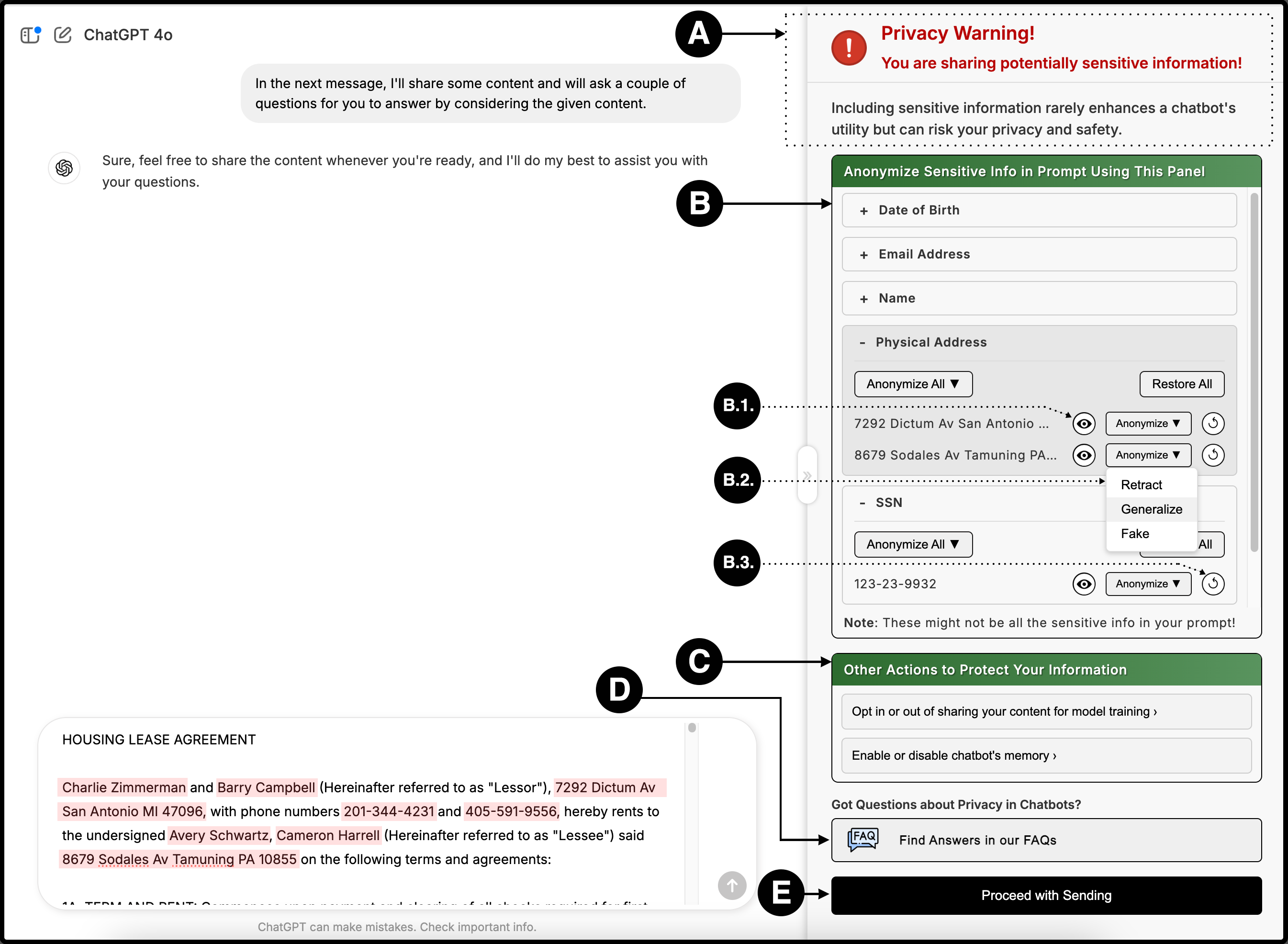}
  \caption{ChatGPT Interface Simulation With Privacy Notice Panel. The panel appears after sending a message containing sensitive info (highlighted in the input). It includes (A) a warning message, (B) an anonymization panel, (C) shortcuts to built-in privacy controls, (D) FAQs, and (E) a proceed with sending button. The anonymization panel further includes, for each detected instance, a (B.1.) locate icon, (B.2.) drop-down menu of anonymization options, and (B.3.) restore button.}
  \label{fig:with_panel}
\end{figure*}

We built a simulation of ChatGPT’s interface (as of Fall 2024) (Figure \ref{fig:with_panel}), including a message input, conversation panel, and a profile icon that provides access to a settings menu mirroring ChatGPT’s design and allowing users to opt in/out of content sharing for model training and to toggle the chatbot’s memory (Appendix \ref{appendix_settings_panel}). 
When memory is enabled, the simulation considers the last seven messages; when disabled, it considers only the most recent message. 
Although this design does not replicate ChatGPT's exact memory mechanism, it still creates a trade-off between contextual retention and potential privacy protection. 
The simulation sends messages to the \textit{GPT-4o} API and displays the resulting responses.

Adding to this interface, we developed a \textit{privacy notice panel} that intercepts message submissions immediately after users attempt to submit a message (i.e., clicking the submit button) containing any of the following types of potentially sensitive information embedded in the materials used in our user tasks (Section \ref{sec_procedure}): \textit{names of people}, \textit{email addresses}, \textit{phone numbers}, \textit{physical addresses}, \textit{social security numbers}, and \textit{dates of birth}. 
The panel appears on the right side of the screen without blocking the ongoing interaction and its use is optional. 
The panel displays (as annotated in Figure \ref{fig:with_panel}):
\begin{enumerate}[label=\Alph*., itemsep=2pt, topsep=4pt]
    \item \textbf{Warning Message}--informs users that their message contains potentially sensitive information.
    \item \textbf{Anonymization Panel}--lists types of sensitive information detected, each expandable to show additional details. For each type, individual instances are displayed, and users can: \textit{(B.1) locate and highlight} an instance in their message; \textit{(B.2) anonymize} it using one of three options: \textit{retract} (replace with a type label, e.g., 123-456-7890 $\rightarrow$ [Phone number]), \textit{generalize} (retain coarse details such as US state and ZIP code for addresses, or year for date of birth), or \textit{fake} (replace with a dummy value, e.g., Cameron $\rightarrow$ Arron); and \textit{(B.3) restore} the original value.
    Alternatively, users can use \textit{Anonymize All} and \textit{Restore All} to apply an action to all instances of a given information type. 
    A note at the bottom of the panel reminds users that the list may not capture all sensitive information in their message.
    \item \textbf{Shortcuts to Built-in Privacy Controls}--two buttons providing direct access to ChatGPT's built-in controls for (a) opting in/out of content sharing for model training and (b) enabling or disabling memory (Appendix \ref{appendix_settings_panel}).
    \item \textbf{FAQs}--includes five questions expandable for reading the answer. 
    Each answer is presented in two layers: a brief sentence or heading, with an optional expansion providing a one-paragraph explanation. 
    The questions and answers are meant to provide users additional information on how they can conceptualize sensitive information, view their anonymity when interacting with CAs, balance privacy protection with chatbot utility, understand differences between `generative' CAs and earlier database-driven systems, and consider the shared responsibility for privacy protection between users, the technology, and its providers. 
    \item \textbf{Proceed with Sending}--sends the current message.
\end{enumerate}

\subsection{Study Procedure}
\label{sec_procedure}
Our study consists of five phases. 
To reduce bias, we adopted \textit{incomplete disclosure} strategy by withholding study's privacy goals from participants until after participation.
Our study design and protocol was approved by our institution's IRB. 

\textbf{Pre-Test Survey (Phase 1)}: 
Participants completed a survey with open-ended questions designed to explore initial perceptions of privacy. 
Specifically, we asked participants (a) \textit{what they think sensitive information means}, (b) \textit{how important it is to protect sensitive information when interacting with chatbots like ChatGPT}, (c) \textit{what they can do to protect sensitive information when interacting with ChatGPT}, and to (d) describe \textit{any ChatGPT features that would help them protect sensitive information during interaction}. 
Participants were instructed to respond based on their own thoughts, without searching online or consulting others.

\textbf{Task Sessions (Phases 2 and 3)}:
At least one week after Phase~1, participants attended two remote \textit{Zoom}\footnote{https://www.zoom.com/} sessions, separated by a 1-4 day gap depending on availability.
In the first session, they completed one version of the task assignments (A or B) (see next paragraph) using the chatbot \textit{without} the privacy notice panel; in the second session, they completed the alternate task assignment using the chatbot \textit{with} the privacy notice panel.
In both sessions, participants shared their screen while thinking aloud to verbalize their thoughts. 
The first author facilitated all sessions, beginning with an introduction to the study, tasks, and chatbot, followed by providing think-aloud instructions and asking participants to complete a brief practice task. 
To minimize distractions, the researcher muted himself and disabled his video during tasks.
We also allowed participants to freely direct their own chatbot interactions (e.g., prompting, anonymization, evaluating responses).

\textbf{User Tasks.} 
We created two versions of task assignments (A and B) and alternated them across the two task sessions (Phases 2 and 3) for each participant to counterbalance order and content effects. 
Half of the participants completed Assignment~A in the first session and Assignment~B in the second session, while the order was reversed for the remaining participants (Table~\ref{tab:participants}).
Both assignments included comparable tasks involving text summarization, text classification, searching lengthy content (e.g., contracts), and drafting emails. 
Each assignment consisted of three tasks, each with two steps. 
For each step, participants were provided with task descriptions, required data embedded with sensitive information, and a text entry field for submitting responses.
We designed the tasks to (1) approximate real-world chatbot use scenarios and be relevant to CS undergraduate and master’s students; (2) place participants in common situations involving sensitive information disclosure (e.g., unintentional disclosure when sharing lengthy content with chatbots); and (3) avoid explicit mentions of privacy in task descriptions to minimize bias, while encouraging participants to treat the data as real or as their own. 
A detailed description of the tasks and embedded sensitive information is provided in (Appendix \ref{appendix_user_tasks}).

\textbf{Immediate Post-Test Survey (Phase 4)}:
Within one day after Phase~3, participants completed a survey capturing their reflections on the task sessions. 
We included open-ended questions asking participants to describe chatbot features (including the privacy panel) that supported them in protecting sensitive information during interaction; to share their thoughts about the design of each component of the privacy notice panel (i.e., the anonymization panel, FAQs, and shortcuts to built-in privacy controls); to describe how easy or difficult they found it to use the panel; to suggest ways to improve the panel; and to provide any additional comments.

\textbf{Delayed Post-Test Survey (Phase 5)}:
At least one week after Phase~3, participants completed a survey that repeated the pre-test (Phase~1) questions to examine changes in their privacy perceptions. 
The survey also included questions asking whether and how participation in the study influenced participants' thinking about privacy when using chatbots, as well as inviting any additional comments about their participation.
We administered all surveys through  \textit{Qualtrics}\footnote{https://www.qualtrics.com/}.

\subsection{Participants}
\label{section_participants}
Our work focuses on undergraduate and master's students in a CS program in the US. 
We selected this participant group because students are frequent users of chatbots (particularly ChatGPT) \cite{zhang-2025-exploring,stohr-2024}, young users are more prone to disclosing sensitive information \cite{lappeman-2022-trust}, and CS students are likely future developers of chatbots, making their privacy perceptions especially relevant to examine.
Focusing on a specific participant group also allowed us to tailor the design of tasks, surveys, and interfaces to participants' contexts, thereby strengthening the ecological validity of the study---e.g., by including data analysis tasks (e.g., classification), using documents related to participants' semester schedules designed for CS students, and adapting the language of FAQ content and survey questions to be relevant to participants' context.
Additionally, this population enabled us to control for variation in age and geographic context while allowing diversity in gender and frequency of chatbot use \cite{malki-2025-hoovered,Tifferet-2019,Engstrom-2023,Li-2017-cross-cultural}.

We recruited participants through a poster distributed at a large public university in the northeastern US. 
Interested individuals completed a screening survey collecting background information (e.g., degree level, gender, and ChatGPT use). 
We recruited 11 students, one of whom completed a pilot session that informed refinements to the study protocol. 
The final sample included ten participants, ranging from third-year undergraduate to second-year master’s students, evenly split by (self-reported) gender (five male and five female), with varying frequency of ChatGPT use (e.g., daily or several times per week).
Participants completed all study phases between March and June 2025. 
We paused recruitment after ten participants, as our iterative analysis suggested thematic saturation, with no substantively new themes observable in additional responses or sessions. 
Each participant received a \$70 USD gift card. 
Participant details are summarized in Table~\ref{tab:participants}.

\begin{table*}[t]
  \caption{Participant Details (self-reported) and their Task Assignment Versions. \textbf{Session 1} refers to the task session using the chatbot \textbf{without} the panel, and \textbf{Session 2} refers to the session using the chatbot \textbf{with} the panel.}
  \centering
  \setlength{\tabcolsep}{11pt}
  \label{tab:participants}
  \begin{tabular}{lllllll}
    \toprule
    \textbf{ID} 
    & \textbf{Degree - Year} 
    & \textbf{Gender} 
    & \multicolumn{2}{l}{\textbf{ChatGPT Use}} 
    & \multicolumn{2}{l}{\textbf{Task Assignment Version}} 
    \\
    \cmidrule(lr){4-5}
    \cmidrule(lr){6-7}
    & & & \textbf{Duration} & \textbf{Frequency} & \textbf{Session 1} & \textbf{Session 2} \\
    \toprule    
    P1 & Master - 1st year & Female & ~\textasciitilde2.5 year & Daily & A & B \\
    P2 & Undergrad - 4th year & Female & ~\textasciitilde2.5 years & Daily & B & A \\
    P3 & Master - 1st year & Female & ~\textasciitilde2.5 years & Several times a week & A & B \\
    P4 & Master - 2nd year & Male & ~\textasciitilde2.5 years & Daily & B & A \\
    P5 & Undergrad - 3rd year & Male & ~\textasciitilde1.5 years & Daily & A & B \\
    P6 & Undergrad - 4th year & Male & ~\textasciitilde2 years & Several times a week & B & A \\
    P7 & Undergrad - 4th year & Female & ~\textasciitilde2 years & Several times a week & A & B \\
    P8 & Undergrad - 3rd year & Male & ~\textasciitilde1 year & Once a week & B & A \\
    P9 & Master - 1st year & Male & ~\textasciitilde3 years & Daily & A & B \\
    P10 & Undergrad - 3rd year & Female & ~\textasciitilde1.5 years & Daily & B & A \\
    \bottomrule
  \end{tabular}
\end{table*}

\subsection{Data Collection and Analysis Approach}
\label{sec:analysis}
For each participant, we collected responses to all pre- and post-test surveys along with audio and video recordings of the two task sessions, including think-aloud transcripts. 
Each task session, excluding the introduction and practice task, lasted between 15 and 80 minutes.

We analyzed qualitative data using thematic coding \cite{braun-2006}, with procedures tailored to each research question (RQ).
To examine \textit{RQ1.1}, \textit{RQ1.2}, and \textit{RQ1.3}, we grouped participant responses from the pre-test (Phase 1) and delayed post-test (Phase 5) surveys according to the relevance of each survey question to the corresponding RQ.
For data relevant to \textit{RQ1.1} and \textit{RQ1.2}, we conducted inductive thematic coding to derive codes directly from responses. 
For \textit{RQ1.3}, we employed a hybrid thematic coding (inductive and deductive), initializing the codebook with codes corresponding to privacy-protective actions supported by the chatbot and privacy notice panel (e.g., retracting, faking, disabling memory) and iteratively refining the codebook based on the data.
For each data group, we followed an iterative process in which recurring and related codes were identified and grouped to develop a codebook with representative themes and detailed descriptions. 
Two authors then independently applied each codebook to its associated dataset and met in weekly collaborative interpretation sessions to resolve ambiguities, reconcile disagreements, and finalize theme definitions through shared interpretation of the data. 
The resulting themes are presented in Table~\ref{tab:themes} and described in Section~\ref{sec:results}.
To examine \textit{RQ2}, we inductively coded the think-aloud transcripts from the task sessions (Phases 2 and 3) and responses to the immediate post-test survey (Phase 4). 
While themes were derived inductively from the data, we oriented the analysis toward understanding participants' perspectives on interface and experience design features of our privacy panel that supported or hindered interactions (Section \ref{results_rq2}).

\section{Results}
\label{sec:results}
We present themes from participants' responses concerning their conceptualizations of sensitive information (\textit{RQ1.1}; Section~\ref{results_rq1.1}), views on the importance of protecting it during chatbot use (\textit{RQ1.2}; Section~\ref{results_rq1.2}), and available protective actions (\textit{RQ1.3}; Section~\ref{results_rq1.3}).
These are summarized in Table~\ref{tab:themes}.
For each theme, we provide a brief description followed by illustrative quotes from pre- and post-test responses. 
Participant IDs shown in \textbf{bold} indicate those who articulated a theme after the study but not before.
We then present students' reflections on how participation in the study influenced their privacy thinking (Section~\ref{results_reflections}) and how the design of our privacy panel supported or hindered user-led protection of sensitive information (\textit{RQ2}; Section~\ref{results_rq2}).

\begin{table*}[t]
\small
\centering
\caption{Columns (L$\rightarrow$R): RQ, identified Themes \& Sub-Themes, the list of participants who articulated each theme before and after the study. \textbf{Bold} participant IDs denote those who discussed the theme after the study but not before.}
\label{tab:themes}
\begin{tabular}{p{2.5cm} p{6cm} p{3.4cm} p{4.3cm}}

\toprule
\textbf{RQ} & \textbf{Themes \& Sub-Themes} & \textbf{Before Study} & \textbf{After Study} \\
\midrule

\multirow{5}{2.5cm}{RQ1.1. How students conceptualize sensitive information.} 
 & \textbullet \thinspace Potentials for harm if leaked (\textit{\ref{rq1.1_harm_leaked}}) & P1, P2, P3, P5, P6, P7, P8 & P1, P2, P3, P7, \textbf{P10} \\
 & \textbullet \thinspace Restricted access (\textit{\ref{rq1.1_access}}) & P1, P4, P5, P6, P7, P9, P10 & \textbf{P3}, P4, P5, P6, P7, P9, P10 \\
 & \textbullet \thinspace Linkable to identity (\textit{\ref{rq1.1_linkable}}) & P2, P3, P4, P8, P10 & P2, P3, P8 \\
 & \textbullet \thinspace Context-based sensitivity (\textit{\ref{rq1.1_context}}) & P5, P9 & \textbf{P6}, P9, \textbf{P10} \\
 & \textbullet \thinspace Security measures (\textit{\ref{rq1.1_security}}) & P5 & P6 \\
\midrule

\multirow{5}{2.5cm}{RQ1.2. Views on the importance of protecting sensitive info during chatbot use} 
 & \textbullet \thinspace Unclear data lifecycle (\textit{\ref{rq1.2_lifecycle}}) & P1, P4, P5, P6, P8, P9, P10 & \textbf{P3}, P4, P5, P6, \textbf{P7}, P8, P9 \\
 & \textbullet \thinspace Unauthorized access to data (\textit{\ref{rq1.2_unauthorized}}) & P2, P3, P4, P9, P10 & P2, P3, P4, \textbf{P8}, P9, P10 \\
 & \textbullet \thinspace Potential consequences of data misuse (\textit{\ref{rq1.2_consequences}}) & P1, P3, P6, P10 & \textbf{P7}, P10 \\
 & \textbullet \thinspace Corp. distrust and lack of transparency (\textit{\ref{rq1.2_distrust}}) & P2, P5, P8 & \textbf{P3}, \textbf{P4}, \textbf{P6}, \textbf{P7} \\
 & \textbullet \thinspace Self-disclosure needs (\textit{\ref{rq1.2_selfdisclosure}}) & P1, P7 & - \\

\midrule

\multirow{9}{2.1cm}{RQ1.3. What students think they can do to protect sensitive information during ChatGPT use.} 
 & \textbullet \thinspace Withholding/anonymizing (\textit{\ref{rq1.3_withhold/anonymize}}) &  &  \\
 & \hspace{2.5em}\textbullet \thinspace Information exclusion & P1-10 & P2, P3, P9, P10 \\
 & \hspace{2.5em}\textbullet \thinspace Instance-based masking  & P2, P4, P9 & \textbf{P1}, P2, \textbf{P3}, P4, \textbf{P5}, \textbf{P6}, \textbf{P7}, \textbf{P8}, P9, \textbf{P10} \\
 & \hspace{5em}\textbullet \thinspace Retracting & P4, P9 & \textbf{P1}, \textbf{P2}, \textbf{P3}, P4, \textbf{P5}, \textbf{P6}, \textbf{P7}, \textbf{P8}, P9 \\
 & \hspace{5em}\textbullet \thinspace Faking & P2 & P2, \textbf{P3}, \textbf{P4}, \textbf{P7}, \textbf{P8}, \textbf{P9}, \textbf{P10}  \\
 & \hspace{5em}\textbullet \thinspace Generalizing & - & \textbf{P3}, \textbf{P9} \\
 & \textbullet \thinspace Using built-in privacy controls (\textit{\ref{rq1.3_builtin}}) &  &  \\
 & \hspace{2.5em}\textbullet \thinspace Opting out of sharing content & P4, P9 & P4 \\
 & \hspace{2.5em}\textbullet \thinspace Disabling memory & - & \textbf{P2}, \textbf{P7} \\

\bottomrule
\end{tabular}
\end{table*}

\subsection{How students conceptualize sensitive information (RQ1.1)}
\label{results_rq1.1}
\subsubsection{Potentials for Harm if Leaked}
\label{rq1.1_harm_leaked}
Under this conceptualization, students viewed information as sensitive if access by others could enable harm to information owners or to whom the information pertains.
Before the study, responses from seven students (\emph{P1, P2, P3, P5, P6, P7, P8}) reflected this view---e.g., \emph{P3} noted that the leakage of sensitive information \emph{``leads to severe consequences to the person including identity theft.''} 
Similarly, \emph{P6} explained: \emph{``[...][access to such information] could potentially result in security threats or danger of some kind.''}
In post-test responses, five students (\emph{P1, P2, P3, P7, \textbf{P10}}) showed this conceptualization, including \emph{P10}, who had not articulated it before the study, saying: \emph{``It is a piece of data that we can't share with anybody and its loss can lead to serious crimes such as identity theft, misuse of data and so on.''}

\subsubsection{Restricted Access}
\label{rq1.1_access}
Under this theme, types of info are considered sensitive when students want them to remain confidential or shared only with specific people, rather than accessed publicly through viewing, knowing, or possession.
Before the study, seven students (\emph{P1, P4, P5, P6, P7, P9, P10}) articulated this conceptualization---e.g., \emph{P10} stated: \emph{``It is similar to having FERPA rights and only certain people should have access to that info.''} and \emph{P5} noted: \emph{``[this information] should only be shared with a selected group of people.''}
After the study, we observed this theme in the responses of seven students (\emph{\textbf{P3}, P4, P5, P6, P7, P9, P10}), including \emph{P3} who had not articulated it prior to the study.

\subsubsection{Linkable to Identity}
\label{rq1.1_linkable}
This theme characterizes information as sensitive when it can be used to identify individuals, either directly or indirectly (similar to Personally Identifiable Information, PIIs). 
Five students (\emph{P2, P3, P4, P8, P10}) expressed this view in their pre-test responses---e.g., \emph{P2} emphasized the \emph{traceability} of such information: \emph{``I think sensitive information is any information that can be traced back to find out who it belongs to [...].''}
Similarly, \emph{P3} noted: \emph{``Sensitive information is information that [...] can lead to the identity of an individual.''} 
Post-test responses reflected this theme for three students (\emph{P2, P3, P8}), all of whom had articulated this view prior to the study.

\subsubsection{Context-based Sensitivity}
\label{rq1.1_context}
This theme emphasizes that what is considered sensitive can vary across individuals and context. 
Two students showed this view prior to the study (\emph{P5, P9})---e.g., \emph{P5} described the subjectivity of sensitive information: \emph{``Sensitive info depends on who it is sensitive for [...]''}, and \emph{P9} emphasized its context dependence: \emph{``It depends largely on the context for me.''}
After the study, responses from three students reflected this perspective (\emph{\textbf{P6}, P9, \textbf{P10}}), including \emph{P6} and \emph{P10}, who had not noted it before the study---e.g., \emph{P6} described protections based on owner's preferences \emph{``[...][sensitive info] can be protected depending on what the source/owner of the info wants.''}

\subsubsection{Security Measures}
\label{rq1.1_security}
This theme captures responses that characterize sensitive information as those often protected or secured by specific technical measures.
\emph{P5} articulated this prior to the study, and \emph{\textbf{P6}} talked about it after the study---e.g., \emph{P5} stated: \emph{``This information is usually handled with multiple levels of security to prevent leaks.''}

\subsection{Students' views on the importance of protecting sensitive info in chatbot use (RQ1.2)}
\label{results_rq1.2}
\subsubsection{Unclear Data Lifecycle}
\label{rq1.2_lifecycle}
This theme captures views on the importance of protecting sensitive information during chatbot use, driven by concerns and uncertainties about what happens to user data after submission, including where it is stored, how long it is retained, how it is used in back-end processes such as model training and response generation, and the possibility of users’ data being sold to third parties.
Before the study, seven students (\emph{P1, P4, P5, P6, P8, P9, P10}) shared such concerns. 
For example, \emph{P4} discussed the difficulty of removing sensitive details once it is used to train models: \emph{``[Protecting sensitive info during CA use is] very important, because these providers potentially train on chats, making it very hard to remove the info.''} and \emph{P1} elaborated on ambiguities around data use \emph{``it’s not very clear how this data could be used, unlike platforms like Google, where it's obvious your data might be used for targeted ads or recommendations. With ChatGPT, there’s no visible feedback loop like that, so it doesn't feel like your data is being used in any particular way.''}
After the study, we observed this theme in the responses of seven students (\emph{\textbf{P3}, P4, P5, P6, \textbf{P7}, P8, P9}), including \emph{P3}, and \emph{P7}, who had not mentioned it before the study. 
For instance, \emph{P7} discussed the possibility of training models on private data: \emph{``I know that they could use private data to train on and improve performance [...],''} noting that this view resulted from participating in this study. 

\subsubsection{Unauthorized Access to Data}
\label{rq1.2_unauthorized}
Students described ways in which sensitive information shared with chatbots (like ChatGPT) could be accessed by unauthorized parties to emphasize the importance of protecting sensitive details. 
Before the study, five students (\emph{P2, P3, P4, P9, P10}) articulated such concerns. 
For example, \emph{P9} shared worries that sensitive information could be accessed through targeted prompts, stating: \emph{``Since [sensitive details] are forever stored in the database, any other user interacting with ChatGPT could ask targeted questions to it to maliciously get that information. The chatbot not understanding the sensitive nature of this information might also provide it to the user.''}
Similarly, \emph{P2} raised concerns about unauthorized access through compromised accounts, noting: \emph{``[...] even if the chats are encrypted, if my ChatGPT login credentials are lost, somebody can access it and misuse it.''}
After the study, six students (\emph{P2, P3, P4, \textbf{P8}, P9, P10}) described similar concerns. 
For instance, \emph{P4} discussed the importance of protections when opting in to data sharing for model training: \emph{``If you're opting for data sharing, it's important to protect your sensitive information since it can be used for training and an attacker can prompt the model to share your info.''}

\subsubsection{Potential Consequences of Data Misuse}
\label{rq1.2_consequences}
Students emphasized the importance of protecting information because of the potentials for misuse and its serious consequences.
Pre-test responses from four students (\emph{P1, P3, P6, P10}) reflected such concerns.
For example, \emph{P3} described examples of such potential consequences, stating: \emph{``That kind of information can be used for financial fraud, identity theft, and plagiarism (it already is and is scrutinized heavily in the AI ethics sphere). It can enable cybercrime and make it more prominent by giving easy access to nefarious parties.''}
Similarly, \emph{P1} emphasized the implications that could arise from behavioral analysis using sensitive information, saying: \emph{``Thinking about it more, I do see how this kind of data could be a goldmine for behavioral analysis. These are not just search terms, they are complex conversations that reveal thought patterns, emotions, and personal struggles. That kind of information, if misused, could have serious implications.''}
In post-test responses, two students (\emph{\textbf{P7}, P10}) discussed this theme---e.g., \emph{P7} elaborated on additional risks: \emph{``With the amount of information a chatbot can collect about a person, lots of metadata and user preferences can be accessed, leading to better targeted ads or even possibly surveillance by governmental figures if the data is sold or released to others.''}

\subsubsection{Corporate Distrust and Lack of Transparency}
\label{rq1.2_distrust}
Students find it important to protect sensitive information due to a lack of trust in chatbot owners \cite{castro-2025-casestudies}. 
They attributed this distrust to factors including ambiguity around data handling practices and past privacy violations. 
Before the study, three students (\emph{P2, P5, P8}) expressed such perspectives. 
For example, \emph{P8} raised concerns about uncertainty in how data is handled, stating: \emph{``I don't personally know exactly where that information goes and how transparent OpenAI is about that,''} and \emph{P5} highlighted distrust in large corporations and chatbot owners, noting: \emph{``Although it may be secure (as they say), it feels wrong to give my personal information to a big corporation. Many of which consistently break good privacy practices and sell information to third parties. This is more concerning when realizing that most chatbots are offered for free, which means they have to monetize some other way, either with the paid version or selling information.''}
After the study, four students (\emph{\textbf{P3}, \textbf{P4}, \textbf{P6}, \textbf{P7}}) expressed similar concerns, none of whom had articulated this view prior to the study. 
For instance, \emph{P6} voiced concern that chatbot owners may prioritize profit over users' privacy, stating: \emph{``I don't believe we should be giving these models our information to allow them to become better [...]. We should always avoid giving information to corporations as they never have our best interests at heart.''}

\subsubsection{Self-Disclosure Needs}
\label{rq1.2_selfdisclosure}
Two students (\emph{P1, P7}) emphasized the need for greater caution in protecting sensitive information during CA use, expressing concerns about a tendency (either in themselves or others) to disclose highly sensitive information to CAs that they would not typically share with friends or family, attributing this tendency to needs for self-disclosure (see below). 
None of the participants discussed this theme in their post-test responses.
\begin{quote}
P7: \emph{``When I'm feeling overwhelmed with CS/career stuff, I tend to type a long brain dump to feel like I'm telling someone (even if it's just ChatGPT). I've shared insecurities and fears with it that'd be hard for me to even share with a close friend.''}
\end{quote}
\begin{quote}
P1: \emph{``I’ve noticed that people, including some of my friends and family, tend to overshare very personal things with chatbots, things they probably wouldn’t say out loud or share with a friend. Sometimes they treat it like a therapist or use it to organize their thoughts. I don’t usually do that myself, not because I’m consciously guarding my privacy, but because I’ve never felt the need to use it that way.''}
\end{quote}

\subsection{What students think they can do to protect sensitive info during ChatGPT use (RQ1.3)}
\label{results_rq1.3}

\subsubsection{Withholding or Anonymizing Sensitive Info}
\label{rq1.3_withhold/anonymize}
This theme covers approaches for protecting sensitive information before submitting them to the chatbot. 
Within this theme, we identified the following two sub-themes.

\paragraph{Information Exclusion.}
This strategy involves withholding entire segments of data that users deem sensitive. 
Before the study, \emph{all participants} mentioned this strategy---e.g., \emph{P8} stated: \emph{``Kind of obvious, but don't put it in the chat! Beyond that, I'm not super sure what can be done [...]''}, similarly, \emph{P10} noted: \emph{``I think the best way is not to share any sensitive info when we are interacting with LLMs.''}
After the study, four students (\emph{P2, P3, P9, P10}) described this strategy again.

\paragraph{Instance-Based Anonymization.}
This strategy involves masking instances of sensitive information before submitting them to the chatbot and includes the \textit{Retracting}, \textit{Faking}, and \textit{Generalizing} approaches.
Before the study, only three students (\emph{P2, P4, P9}) described instance-based anonymization: \emph{P4} and \emph{P9} mentioned the \textit{Retracting} strategy, while \emph{P2} described \textit{Faking}---e.g., \emph{P9} noted: \emph{``[...] replace that info with a placeholder so that ChatGPT does not have the exact info.''}
After the study, \textit{\textbf{all students}} discussed instance-based anonymization approaches. 
Specifically, \textit{\textbf{all except P10}} described \textit{Retracting}; \emph{P2, \textbf{P3}, \textbf{P4}, \textbf{P7}, \textbf{P8}, \textbf{P9}, \textbf{P10}} discussed \textit{Faking}; and \emph{\textbf{P3}} and \emph{\textbf{P9}} described \textit{Generalizing}.
For example, \emph{P2} emphasized \textit{Retracting} and \textit{Faking}: \emph{``I can retract the sensitive info or fake the data instead of providing actual sensitive info that belongs to people''}, and \emph{P3} highlighted the \emph{Faking} and \emph{Generalizing} strategies: \emph{``We can provide fake info/generalized info that doesn't pinpoint to an individual.''}

\subsubsection{Using ChatGPT's Built-in Privacy Controls}
\label{rq1.3_builtin}
This theme captures the use of features or settings built into \textit{ChatGPT} to help protect sensitive information after submission. 
Before the study, three students mentioned using \textit{temporary chats} (\emph{P1, P4, P9}), two mentioned \textit{removing chat histories} (\emph{P1, P5}), and two mentioned \textit{opting out of sharing content for model training} (\emph{P4, P9}). 
After the study (focusing only on approaches supported by our simulated chatbot), two students (\emph{\textbf{P2}, \textbf{P7}}) mentioned \textit{disabling the chatbot’s memory}, and one student (\emph{P4}) mentioned \textit{opting out of content sharing}.

\subsection{Students' Reflections on Changes in Their Privacy Thinking}
\label{results_reflections}
\subsubsection{Desire for User-Facing Privacy Tools in CAs}
This theme captures participants' interest in user-facing tools or features that support them in protecting sensitive information during chatbot interactions (\emph{P1, P2, P3, P4, P5, P6, P8, P9}). 
For example, \emph{P6} emphasized that such features can help balance privacy with chatbot usefulness, noting: \emph{``By having a UI feature that identifies and hides all our sensitive information given our input, it's super useful to make sure the model is useful while still protecting ourselves.''} 
Similarly, \emph{P5} discussed the practicality of such tools for their own chatbot use, particularly if they are chatbot-agnostic: \emph{``I think it was interesting to have a tool dedicated to censoring sensitive information. I think if it existed in real life, I would be inclined to use it. However, it would have to either be chatbot-agnostic, meaning that it could be used with any chatbot, or it would have to be a chatbot I already use.''}

\subsubsection{Growing Privacy Awareness and Thinking}
Participants explained that participating in the study and interacting with the privacy panel prompted them to think more about privacy and become more cautious about the data they submit to chatbots (\emph{P2, P3, P5, P7, P8, P9, P10}). 
For example, \emph{P2} reflected on how their disclosure behavior may be changed after the study: \emph{``I've used ChatGPT to write replies for emails or summarize different things before the study as well without really checking if it contains any private information. But after the study, I have started to think more before sending my prompts and removing or faking such data wherever possible/necessary.''}
Similarly, \emph{P5} described how interacting with the privacy panel during the task sessions made them more aware of how much data they had been sharing with chatbots: \emph{``[...] it has opened my eyes to how much data I was feeding the chatbots without knowing it. I had never really thought of it before, but when I actually was tasked with censoring that data, I realized just how much of it I might've been leaking.''}

\subsection{Interface Design Features That Supported or Hindered Protection of Sensitive Information (RQ2)}
\label{results_rq2}

\subsubsection{Intercepting Chatbot Interactions After Each Submission of Sensitive Information}
\label{sec_justintime_design}
We designed the privacy panel to automatically intercept interactions immediately after users attempt to send a message containing sensitive information and before it is sent to the API. 
Participants generally found this interception helpful, as it made the panel easy to access, intuitive to use, and minimally disruptive to their workflow (\emph{P2, P3, P4, P5, P6, P7, P9}). 
For example, \emph{P2} shared that using the panel was easy because: \emph{``it automatically popped up every time there was any personal information in the prompt so it was intuitive and easy to access.''}
\emph{P6} and \emph{P9} emphasized the benefit of this design on raising awareness of sensitive content in their messages at moments when they could make context-based decisions (e.g., based on the given task) about how to handle the information.
\emph{P7} described the benefit of integrating the panel into the existing workflow (e.g., avoiding navigation to other screens), stating: \emph{``If this was a feature I had to go out of my way to interact with, I probably wouldn't think about what private information I might be sharing, and would just send the message straight away.''}
\emph{P7} also suggested improvements for situations in which users resize their browser window or use a split-screen view, noting that the panel may require automatic resizing to enable simultaneous use of both the panel and the chatbot.
Despite these benefits, \emph{P5} and \emph{P9} found the repeated interceptions sometimes tedious, especially when they did not consider the flagged information as truly sensitive, with \emph{P5} saying: \emph{``[...] there were a few things that made it more annoying, like how it popped up every time, even though [sharing] names aren't as much of a privacy risk personally.''}

\subsubsection{Interface Terminology and Features for Understanding the Panel}
\label{sec_terminology}
The panel used \textit{Anonymizing} as an umbrella term for three strategies labeled as \textit{Retracting}, \textit{Faking}, and \textit{Generalizing} (Figure \ref{fig:with_panel}).
Four participants (\emph{P4, P6, P7, P9}) faced difficulty in distinguishing how the options work just by their labels, as \emph{P6} noted: \emph{``I wasn't sure about the exact differences between anonymizing vs. retracting vs. generalizing vs. faking.''}
Among them, three (\emph{P4, P7, P9}) shared that the panel's \emph{locate and highlight} feature helped understand how each strategy worked.
\emph{P7} explains this: \emph{``The feature to jump to the part of the text being anonymized/faked was very helpful [...] as it allowed me to follow what was being changed and how it was being changed according to which option I selected.''}
Despite its usefulness, the \textit{locate and highlight} feature appeared to lack visibility as it was unused by four participants (\emph{P1, P2, P3, P6}---excluding \emph{P8} and \emph{P10} who used the panel minimally). 
\emph{P6}, unaware of the feature's presence, suggested adding such functionality to the panel: \emph{``Maybe an example being highlighted to show what was being edited in the text could be helpful?''} 
We also noticed that three participants (\emph{P2, P6, P9}) initially thought the panel automatically anonymized flagged information, indicating a need for clearer status signifiers.
For example, \emph{P2} misunderstood the “Anonymize all” label and proposed renaming it to reduce confusion: \emph{``[...] I assumed it's already done but later when I clicked it, I saw the options to fake or generalize or [retract] so, maybe like [using] `Select an option' default value.''}

\subsubsection{Quick and Low-Effort User Control Over Anonymization}
\label{sec_quick_effort}
Participants frequently appreciated features of the panel that supported quick and low-effort anonymization of sensitive information (\emph{P1, P3, P5, P9}), as \emph{P1} noted: \emph{``I liked the feature that allowed me to quickly retract private information like name, email, phone number.''}
They valued design elements that enabled their control over anonymization decisions, including the ability to use the panel optionally (\emph{P5}), view all detected instances and apply anonymization actions individually or in bulk (\emph{P2, P4, P6, P9}), and track or restore changes as needed (\emph{P4, P9}). For example, \emph{P4} shared: \emph{``I thought it was very helpful to show every instance of PII and adding an option to anonymize all, as well as restore all.''}
Some participants also suggested ways to further speed up the process, such as saving preferred anonymization actions for future prompts (\emph{P5, P9}) or adding a universal “Anonymize All” button to the panel that applies to all types and instances of information (\emph{P9}).
These reflections highlight a core challenge in interface design: balancing user control with efficiency, especially when aiming to support informed and context-based privacy actions with minimal effort.

\section{Discussion and Future Work}
\label{sec:discussion}
\subsection{In-Context Support and Experiential Privacy Learning}
\label{sec_dis_experiential}
Although pre-test responses showed that participants had some understanding of sensitive information, recognized the importance of protecting it during CA use, and were aware of several protective strategies, 8 of 10 showed no evidence of considering sensitive information protection during the first task session (when using the chatbot \emph{without} the panel) and disclosed most embedded sensitive information. 
This pattern reflects the \emph{privacy paradox} \cite{Kokolakis-2017}, where users' stated privacy attitudes do not align with their actual behaviors.
In contrast, during the second task session (using the chatbot \emph{with} the panel), participants began actively reasoning about what to protect and how, withholding or anonymizing most sensitive information before submission.
The differences between participant behaviors across sessions suggests that knowledge (e.g., what is sensitive, protective actions) and motivation (e.g., valuing privacy protection) may not be sufficient by themselves to lead users to taking protective actions. 
Instead, design features that facilitate in-context protective actions appear to encourage greater engagement with privacy-preserving practices.
These elements (knowledge, incentives, and in-context action) are actually closely related to \emph{experiential learning}, or \emph{learning by doing} \cite{illeris-2007, kolb-2014} (see also models of sensemaking and decision-making \cite{wise-2019-teaching,hadi-2024-math})
Learning involves interpreting new information through the lens of prior knowledge \cite{ambrose-2010-learning}, and experiential learning emphasizes the importance of exposure to new content, incentives for engagement, and opportunities for interaction (action and reflection) \cite{illeris-2007}. 
Below, we use this lens to interpret the potential impact of our panel components on changes in participants' perspectives from pre- to post-test responses.

\textbf{Anonymization Component (Figure \ref{fig:with_panel}, B)}: 
In the second task session, most participants engaged with the anonymization component, and post-study responses suggest that they learned about instance-based masking strategies (e.g., retract, fake). 
This component introduced potentially new strategies, supported users in applying them, and may have prompted reflection on their effects---e.g., by considering changes in chatbot response quality. 
These supports were embedded within ongoing interactions, appearing right after each prompt submission, allowing participants to consider contextual factors (e.g., tasks and goals) when making decisions. 
Integrating these features into the interaction flow may have facilitated learning about these privacy-protective strategies.

\textbf{FAQ Component (Figure \ref{fig:with_panel}, D)}:
This component provided content related to conceptualizing sensitive information and the importance of privacy protection. 
During the second task session, only two participants opened the FAQ panel, and only one read parts of it. 
Not surprisingly, we observed minimal changes in participants' perspectives on sensitive information and the importance of privacy protection in the post-study responses.
One possible explanation is that participants may have perceived the FAQs as static, text-heavy resources that were less directly integrated into or actionable within their immediate interactions. 
Future work could explore how informational components can be more tightly embedded in interaction flows and designed to support action. 
For example, the anonymization component (Figure~\ref{fig:with_panel}, B) could include brief contextual prompts explaining why flagged information may be sensitive and why protecting it matters at that moment.

\subsection{UI/UX Design Decisions for Engagement and Learning in User-Facing Privacy Tools}
\label{sec_dis_uiux}
\textbf{Interface Terminology and Language}:
In our study, although some participants found it difficult to distinguish between the terms used for different instance-based masking strategies (i.e., anonymize, retract, fake, generalize), they were able to understand how each strategy worked through interaction with the panel---by applying a strategy, locating, highlighting, and undoing changes (Section~\ref{sec_terminology}). 
This suggests that interactive features can help compensate for unfamiliar or complex terminology.
An alternative approach is to incorporate contextual on-boarding techniques \cite{nngroup-onboarding}---e.g., brief tooltip explanations shown on hover, or in-context pop-ups during initial use, could clarify new terms and features. 
Such approaches may support gradual, in-context learning and encourage engagement.
Regarding the FAQ component (Figure~\ref{fig:with_panel}, D), the limited engagement we observed may relate to how the label ``FAQ'' was interpreted. 
It may have signaled instructions about using the panel rather than content about privacy, as \emph{P2} noted: \emph{``I think the other features [of the panel] were pretty self-explanatory so, I didn't feel the need to read the details in the FAQs.''}

\textbf{Prioritizing Features and Content}: 
User engagement with interface elements can be influenced by how they are prioritized within the design. 
Users typically scan panels from top to bottom \cite{BYRNE-2001}, and our participants followed this pattern: all read the warning message first, almost all engaged with the anonymization panel, some reviewed built-in privacy controls, and only two opened the FAQs (Figure~\ref{fig:with_panel}). 
This ordering likely contributed to limited engagement with FAQs, especially given users' time and effort constraints.
Engagement may also depend on whether content appears in the primary interaction layer or is nested in secondary layers (e.g., additional screens). 
Our participants appreciated that the panel was embedded within their workflow without requiring navigation to another page (Section~\ref{sec_justintime_design}); however, accessing the FAQs required opening a secondary pop-up, which may have further discouraged interaction.
Future work can therefore examine how to prioritize and surface privacy tool components, placing higher-priority features in the primary interaction layer and reducing reliance on secondary screens or nested content (e.g., through \textit{progressive disclosure} \cite{IxDF-progressive-disclosure,nngroup-progressive-disclosure}).

\subsection{Privacy Tradeoffs in CA Use}
\label{sec_dis_tradeoffs}
In CA use, users weigh the costs and benefits of their actions against their goals. 
Advancing one objective may require compromising another. 
Recognizing these trade-offs is essential for supporting users in balancing privacy with other chatbot use objectives. 
Below, we discuss two such trade-offs.

\textbf{Privacy Protection versus Chatbot Utility}:
Users often weigh whether (and how much) to withhold information to protect privacy while still providing sufficient context for high-quality chatbot responses (utility) \cite{zhang-2024,Zhou-2025-rescriber,mutahar-2025-understanding}. 
How users navigate this tradeoff may depend on the perceived importance of each objective, their knowledge of potential actions, and the support available to carry them out.
Without privacy features, users may prioritize chatbot utility over privacy, as observed in our first task session. 
Pre-test responses further suggest that participants were largely unaware of privacy-protective actions beyond fully excluding segments of information (Table \ref{tab:themes}). 
However, such exclusion may be perceived as substantially reducing response quality and therefore avoided.
When the panel introduced and supported instance-based masking strategies (which may have less impact on response quality) participants adopted these strategies. 
Nevertheless, some continued to share sensitive details flagged by the panel, often citing their importance for the chatbot to generate high-quality responses.
This suggests that future work should explore tools that better support users in balancing privacy and chatbot utility---e.g., tools that can analyze the ongoing task (e.g., drafting emails) and provide guidance on what to share or withhold to preserve privacy while maintaining utility.

\textbf{Privacy Protection versus Convenience}:
UI/UX research has long emphasized the importance of efficiency (minimizing users' time and effort) in shaping tool adoption and engagement \cite{davis-1989-perceived,nngroup-heuristics}. 
This principle also applies to privacy tools, where convenience becomes an additional objective alongside privacy and chatbot utility \cite{zhang-2024}. 
Although our panel facilitated protective actions (by detecting sensitive info, providing just-in-time awareness, and bulk anonymization), participants still expressed interest in features that would further reduce time and effort (Section~\ref{sec_quick_effort}).
One direction is exploring automation in anonymization based on user preferences---e.g., automatically applying specific masking strategies to certain types of information. 
In our panel, shortcuts to built-in privacy controls appeared each time the panel was displayed (Figure \ref{fig:with_panel}, C).
While initially surfacing these controls promotes awareness, repeatedly displaying static elements may be unnecessary, as we observed some participants reopened the settings panel across multiple appearances simply to confirm that nothing had changed (e.g., that the opt-out toggle remained disabled). 
Components that do not dynamically change (unlike the anonymization component; Figure \ref{fig:with_panel}, B) can therefore be minimized or hidden in later appearances to reduce redundancy and effort.

\subsection{Learning Opportunities for Students}
\label{sec_dis_learning_opp}
Our findings on students' perceptions of privacy in CAs (Sections \ref{results_rq1.1}, \ref{results_rq1.2}, and \ref{results_rq1.3}) can inform opportunities to update students' understanding of privacy in light of evolving AI capabilities.
For example, although identifiable information is widely recognized as sensitive \cite{Belen-Saglam-2022,Song-2025}, only five participants explicitly articulated this in their responses. 
Moreover, this understanding may require further development, as modern AI tools (including generative CAs) can intensify identifiability risks through rapid, real-time aggregation of data from multiple sources \cite{lee-2024-deepfakes}.
Similarly, users' views on the importance of privacy protection may benefit from greater awareness of risks introduced or exacerbated by modern AI systems (e.g., surveillance, intrusion, phrenology) \cite{lee-2024-deepfakes}. 
Standalone educational tools could also support deeper examination of users' privacy thinking, increasing awareness of past privacy incidents, and learning about protective behaviors \cite{feffer-2023,AIAAIC-bard,AIAAIC-LeeLuda,hadi-2025-embedding}.

\section{Limitations}
\label{sec:limitation}
We focused on sensitive information definition, the importance of protecting privacy during CA use, and protective actions. 
Other relevant dimensions (e.g., understanding of CA privacy policies, awareness of past privacy incidents) were beyond the scope of this study. 
Future work could examine these and explore how they may be enhanced through interactions with privacy tools.
Our participants were CS undergraduate and master's students in the US. 
While this focus enabled a more controlled examination within a specific population, we do not intend to generalize our findings to other user groups (e.g., different ages or geographic contexts). 
Future research can replicate and extend this work with broader and more diverse populations to capture a wider range of privacy perspectives and design needs.
Finally, we examined short-term interactions with our privacy panel. 
Longitudinal studies are needed to assess the long-term effects of engaging with such tools during everyday CA use and to understand how repeated exposure may shape users' privacy perceptions over time.

\section{Conclusion}
We examined how exposure to our just-in-time privacy notice panel during realistic chatbot use influenced participants' perceptions of privacy in CAs. 
We also analyzed the UI/UX design features of the panel that supported or hindered user-led protection of sensitive information during chatbot use. 
Our findings suggest that user-facing privacy tools have potentials in gradually and contextually encouraging users to engage with, reflect on, and learn about privacy during CA interactions. 
We further highlighted the critical role of UI/UX design in facilitating such engagement and the importance of accounting for users' privacy tradeoffs when designing privacy tools for CAs.

\section*{Acknowledgments}

\bibliographystyle{plain}
\bibliography{references.bib}

\appendix
\section{Supplementary Methodological Materials}

\subsection{ChatGPT Interface Simulation}
\label{appendix_settings_panel}
\begin{figure*}[t]
  \centering
  \includegraphics[width=\linewidth]{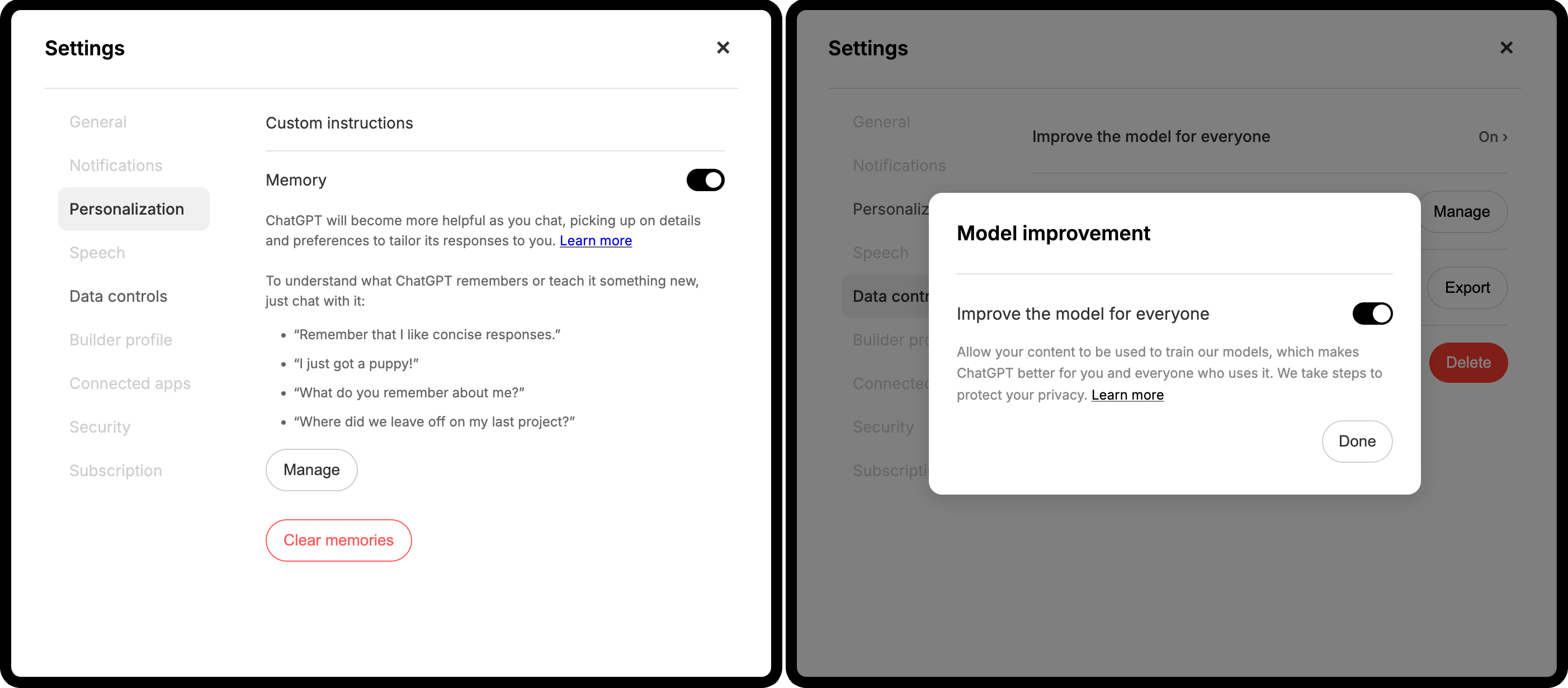}
  \caption{Simulated ChatGPT settings panels for built-in privacy controls: (left) Personalization panel for enabling or disabling memory, with a functional toggle and static buttons; (right) Data Control panel for opting in or out of content sharing for model training, with an interactive toggle.}
  \label{fig:settings-panel}
\end{figure*}

Figure \ref{fig:settings-panel} shows the simulated settings panel for built-in privacy controls: (left) disabling memory and (right) opting out of sharing content for model training. These panels were available in both task sessions.

\subsection{User Task Details}
\label{appendix_user_tasks}
Table \ref{tab:task-overview} shows detailed description of user tasks and the embedded sensitive information including their types, frequencies, and subjects---i.e., who the information pertains or belongs to.

\setlength{\tabcolsep}{6.5pt}
\begin{table*}
    \centering
  \caption{User Tasks. Columns (L$\rightarrow$R): task assignment version (V), task description, types of sensitive info contained in each task (Sensitive Info), frequency---of occurrences---of each type (F), and Info Subjects (who the info pertains/belongs to).}
  \label{tab:task-overview}
  \begin{tabular}{p{0.2cm}p{9cm}p{2.5cm}p{0.3cm}p{3cm}}
    \toprule
    \textbf{V} & \textbf{Task Description} & \textbf{Sensitive Info} & \textbf{F} & \textbf{Info Subject} \\
    \toprule
    A 
    & 
    \textbf{Task 1:} Email Analysis
    \begin{itemize}[label=\textbullet,leftmargin=*,itemsep=0pt,topsep=0pt,parsep=0pt,partopsep=0pt]
        \item \textbf{Step 1:} Classifying seven emails\footnotemark[1] by their sentiments
        \item \textbf{Step 2:} Summarizing two separate email threads
    \end{itemize}
    & 
    Name\newline
    Email address\newline
    Phone number\newline
    Physical address
    &
    61\newline
    23\newline
    6\newline
    4
    & 
    Others (e.g., employees)
    \\
    \midrule
    A 
    & 
    \textbf{Task 2:} Searching Housing Documents
    \begin{itemize}[label=\textbullet, leftmargin=*,itemsep=0pt,topsep=0pt,parsep=0pt,partopsep=0pt]
        \item \textbf{Step 1:} Finding party restrictions from their housing contract (9-page doc)
        \item \textbf{Step 2:} Finding maintenance information from the same housing contract and a welcome letter (5-page doc)
    \end{itemize}
    & 
    Name\newline
    Email address\newline
    Phone number\newline
    Physical address\newline
    SSN\newline
    Date of birth
    & 
    38\newline
    9\newline
    16\newline
    16\newline
    4\newline
    4
    &
    Self, Others (e.g., friend, home owner)
    \\
    \midrule
    A 
    & 
    \textbf{Task 3:} Planning a Trip
    \begin{itemize}[label=\textbullet, leftmargin=*,itemsep=0pt,topsep=0pt,parsep=0pt,partopsep=0pt]
        \item \textbf{Step 1:} Finding timing conflicts between their travel itinerary (2-page doc) and semester schedule (1-page doc)
        \item \textbf{Step 2:} Drafting an email to resolve the conflicts and provide their contact details for reservation purposes
    \end{itemize}
    & 
    Name\newline
    Email address\newline
    Phone number\newline
    Physical address\newline
    Date of birth
    & 
    5\newline
    1\newline
    3\newline
    7\newline
    2
    &
    Self, Others (e.g., friend)
    \\
    \midrule
    B 
    & 
    \textbf{Task 1:} Customer Complaint Analysis
    \begin{itemize}[label=\textbullet, leftmargin=*,itemsep=0pt,topsep=0pt,parsep=0pt,partopsep=0pt]
        \item \textbf{Step 1:} Classifying seven customer complaints\footnotemark[2] by relevant departments
        \item \textbf{Step 2:} Summarizing recommendations for the company from two separate groups of customer complaints
    \end{itemize}
    & 
    Name\newline
    Email address\newline
    Phone number\newline
    Physical address\newline
    Date of birth
    & 
    18\newline
    6\newline
    4\newline
    4\newline
    2
    &
    Others (e.g., customers)
    \\
    \midrule
    B 
    & 
    \textbf{Task 2:} Searching Car \& Housing Documents
    \begin{itemize}[label=\textbullet, leftmargin=*,itemsep=0pt,topsep=0pt,parsep=0pt,partopsep=0pt]
        \item \textbf{Step 1:} Finding costs and fees related to purchasing a leased car from their car lease contract (15-page doc)
        \item \textbf{Step 2:} Finding information about utilities and responsibilities from their housing contract (9-page doc)
    \end{itemize}
    & 
    Name\newline
    Email address\newline
    Phone number\newline
    Physical address\newline
    SSN\newline
    Date of birth
    & 
    23\newline
    4\newline
    10\newline
    12\newline
    4\newline
    4
    &
    Self, Others (e.g., friend, home owner)
    \\
    \midrule
    B 
    & 
    \textbf{Task 3:} Planning Teeth Removal
    \begin{itemize}[label=\textbullet, leftmargin=*,itemsep=0pt,topsep=0pt,parsep=0pt,partopsep=0pt]
        \item \textbf{Step 1:} Finding out-of-pocket expenses from their insurance document (6 pages) and an email from their doctor.
        \item \textbf{Step 2:} Drafting an email to their doctor providing insurance details and asking if the insurance is accepted
    \end{itemize}
    & 
    Name\newline
    Email address\newline
    Phone number\newline
    Physical address\newline
    Date of birth
    & 
    8\newline
    2\newline
    2\newline
    1\newline
    5
    &
    Self, Others (e.g., doctor)
    \\
    \bottomrule
  \end{tabular}
\end{table*}
\footnotetext[1]{Drawn from the \href{http://www.enron-mail.com/email/}{\emph{Enron Email Corpus}} and modified as needed.}
\footnotetext[2]{Drawn from a \href{https://www.kaggle.com/datasets/venkatasubramanian/automatic-ticket-classification}{\emph{Ticket Classification Dataset}} and modified as needed.}

\end{document}